\documentclass[two column,pra,showpacs,letterpaper]{revtex4}

\usepackage{amsmath}
\usepackage{graphicx}
\usepackage{dcolumn}
\usepackage{bm}
\usepackage[utf8]{inputenc}
\usepackage[english]{babel} 
\usepackage[colorlinks, citecolor=red, linkcolor=blue]{hyperref}
\usepackage[section]{placeins} 

\graphicspath{{./Figures/}}

\begin{document}

\preprint{}

\title{Faraday Waves in strongly interacting superfluids}

\author{D. Hernández-Rajkov$^1$}
\author{J. E. Padilla-Castillo$^1$}
\author{A. del Río-Lima$^1$}
\author{A. Gutiérrez-Valdés$^1$}
\author{F. J. Poveda-Cuevas$^2$}
\author{J. A. Seman$^1$}\email{seman@fisica.unam.mx}
\affiliation{$^1$Instituto de Física, Universidad Nacional Autónoma de México, 01000 Ciudad de México, Mexico.\\ $^2$ Cátedras CONACyT - Instituto de Física, Universidad Nacional Autónoma de México, 01000 Ciudad de México, Mexico.}

\begin{abstract}

We report on the observation of Faraday waves in a cigar-shaped Fermi superfluid of $^6$Li parametrically excited by modulating the radial trap frequency. We characterize the phenomenon as a function of the interaction parameter by means of a Feshbach resonance. Starting from the BEC side of the resonance we observe a drop on the visibility of the Faraday pattern as we approach to unitarity, possibly due to the increased incompressibility of the system. We probe the superfluid excitation spectrum by extracting an effective 1D speed of sound for different values of the interaction parameter, in good agreement with numerical simulations. Finally, we perform a stability analysis in the parameter space using a  simplified model and we show the emergence of the Faraday waves as unstable solutions to a Mathieu-like equation.

\end{abstract}


\maketitle

\section{Introduction}\label{sec:intro}

Faraday waves are a non-linear parametric excitation in continuous media that manifest as a spatial and temporal periodic modulation of the density on a non-linear fluid \cite{Cross}. Its study dates back to the 19th century when Michael Faraday first observed them on vertically shaken fluids \cite{Faraday}. Faraday waves are, in fact, a ubiquitous phenomenon in non-linear fluids and have been studied in a large variety of systems such as viscoelastic fluids \cite{Wagner}, granular media \cite{Martino}, and even in living soft systems such as earth-worms \cite{Maksymov}. Naturally, superfluids are not an exception. Faraday waves were first observed and characterized in a weakly interacting BEC by Engels et al. \cite{Engels}, demonstrating its presence in quantum fluids. Since then, they have been studied in these systems from very different perspectives. For instance, Faraday-like patterns were produced in a nearly ideal BEC as a starting point to explore more complex parametric excitations such as granulation \cite{Nguyen}. Smits et al. \cite{Smits}, in turn, identified that Faraday patterns in a BEC present the same behavior as a discrete space-time crystal \cite{Wilczek}, providing a new scenario for the research of nonequilibrium phase transitions and spontaneus symmetry breaking \cite{Smits21}. Moreover, these excitations have also been explored in superfluid liquid $^4$He, where they are a precursor to the formation of classical (non-quantized) vortices \cite{Levchenko}.

From the theoretical point of view, Faraday patterns have also been intensively studied. Several works demonstrate that Faraday waves in BECs arise as solutions of the time dependent Gross-Pitaevskii equation. Indeed, a Mathieu-like equation describing the phenomena can be deduced from this equation \cite{Staliunas2002, Staliunas2004, Nicolin2007, Nicolin2011} and as a result from variational analysis \cite{Nicolin2012}. They have also been investigated in Bose-Bose \cite{Balaz} and Fermi-Bose \cite{Abdullaev} mixtures in which the intra-species scattering length is modulated.

The theory of Faraday waves in Fermi superfluids has also been explored, particularly at the BEC-BCS crossover \cite{Capuzzi, Tang}. However, these excitations have not been experimentally observed in such systems nor in any strongly interacting superfluid gases. Here, we explore these excitations in strongly interacting systems composed by a molecular Fermi superfluid. The advantage of using ultracold Fermi systems over Bose gases is the possibility to tune the interatomic interactions into strongly interacting regimes using a broad Feshbach resonance. 

This article is organized in the following way. In Section~\ref{sec:setup}, we present the experimental setup and methods we use to create Faraday waves. Next, in Section~\ref{sec:fw} we introduce our experimental results together with GPE simulations, exhibiting very good agreement. Later, in Sec.~\ref{sec:strong} we explore the strongest interacting regimes and discuss why our protocol fails to produce the Faraday instability along the BEC-BCS crossover. Finally, in Section~\ref{sec:floquet} we develop an analytic model, based on Floquet theory, which is in good agreement with the data. In particular, this model allows to perform a stability analysis of the phenomenon.

\section{Experimental setup and methods}\label{sec:setup}

The setup and methods employed to produce ultracold samples are described in detail in reference \cite{Rajkov}. We are able to produce ultracold Fermi superfluids formed by a two-component spin mixture of $^6$Li atoms in the two lowest hyperfine states $|F = 1/2, m_F = \pm1/2\rangle $. In this system we can access different superfluid regimes across the BEC-BCS crossover by means of a Feshbach resonance.

Quantum degeneracy is achieved by runaway evaporation in a trap composed by the superposition of a single-beam far red-detuned optical dipole trap (ODT), which tightly confines the atoms along the radial direction, and a magnetic curvature generated by the Feshbach coils, which confines the atoms along the axial direction. At the end of the evaporation we produce samples containing $5 \times 10^4$ pairs in a cylindrical symmetric harmonic trap with frequencies $\omega_r = 2\pi \times 163$~Hz and $\omega_z = 2\pi \times 11$~Hz. The cylindrical symmetry implies that $\omega_x = \omega_y \equiv \omega_r$, we guarantee this condition within a 1\% by finely tuning $\omega_y$ using the time-averaged potential technique described in reference \cite{Roy}.


In all experiments we reach a temperature of $T/T_F = 0.1$, where $T_F$ is the Fermi temperature, corresponding to a condensed fraction above 90\% on the deep BEC regime. The interaction parameter $1/k_Fa_s$ ranges between 0 and 11, where $k_F$ is the Fermi wavevector and $a_s$ is the interatomic scattering length, however, as discussed in Section~\ref{sec:strong}, the Faraday pattern is only visible for $1/k_Fa_s > 2.2$.

In our trap configuration $\omega_r$ and $\omega_z$ are nearly decoupled: the radial frequency depends on the power $P$ of the ODT as $\omega_r \propto \sqrt{P}$, while the axial one depends on the curvature of the magnetic Feshbach field at the sample location $B_z^{''}(0)$ as $\omega_z \propto \sqrt{B_z^{''}(0)}$. In this way, we can manipulate $\omega_r$ independently of $\omega_z$ by varying the ODT power. To excite the superfluid, we modulate the ODT intensity by means of an acousto-optic modulator (AOM) \cite{Rajkov}, this allows us to control the radial frequency profile over time. All excitations studied in this paper are generated by periodically modulating the power of the ODT beam in the form $P(t) = P_0 \left[ 1 + \alpha\sin(\Omega t) \right]$, where $\Omega$ is the frequency of the excitation and $\alpha P_0$ its amplitude. Consequently, the radial trap frequency is  $\omega_r(t) = \omega_r^0 \sqrt{1 + \alpha\sin(\Omega t) }$.

After producing the ultracold sample, we wait an equilibration time of 50~ms and afterwards we apply the following excitation protocol. We modulate the ODT power during a fixed time of ten cycles, $\tau_e = 10/(\Omega/2\pi)$. The amplitude parameter $\alpha$ remains  an important variable that we explore in our experiments. Finally, we image the sample in-situ after a variable evolution time $t$ using an absorption imaging setup with a resolution of  $\sim 2\,\mu$m. Here, $t=0$ corresponds to the moment in which the excitation starts. As we explain in section~\ref{sec:fw}, the excitation frequency is fixed on the radial breathing mode frequency which, for an elongated superfluid in the molecular BEC regime, corresponds to twice the static radial frequency \cite{Hu, Kinast}, i.e. $\Omega = 2\omega_r^0 = 2\pi \times 326$~Hz, hence $\tau_e \simeq 30.7$~ms.

\section{Observation and characterization of Faraday waves}\label{sec:fw}

\begin{figure}
\centering
 \includegraphics[width=\columnwidth]{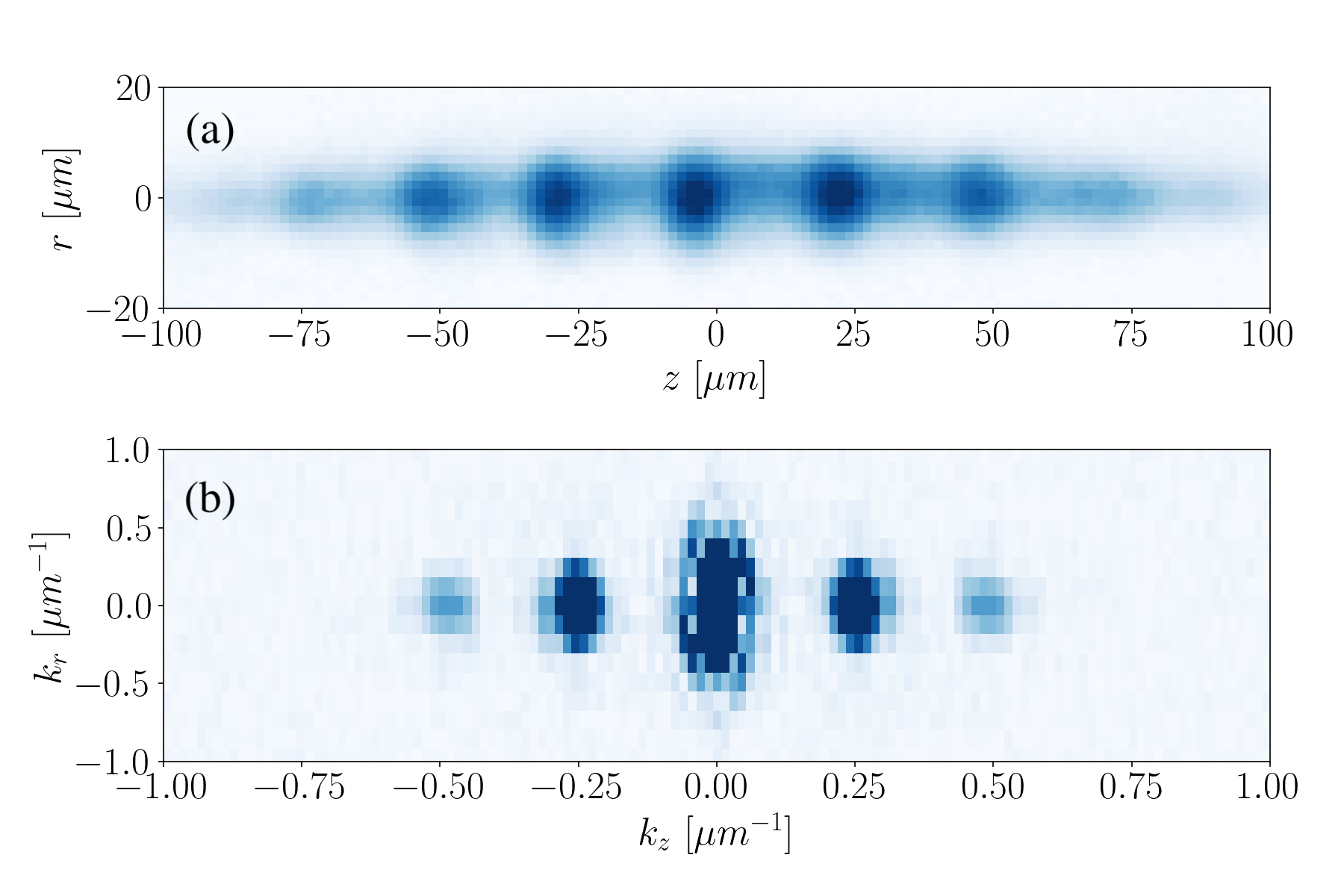}
 \caption{(a) In-trap absorption images of Faraday waves in the superfluid. (b) Fourier transform of the absorption image in (a). Image taken at a magnetic field of 690~G, corresponding to $1/k_Fa_s = 7.1$. Similar images are obtained for other values of $1/k_Fa_s$. }
\label{fig:fig1}
\end{figure}

Once the excitation is applied, we observe the breathing mode with its characteristic frequency $\omega_b = 2\, \omega_r^0$ for an elongated BEC \cite{Hu, Kinast}, independently on the value of the studied $1/k_Fa_s$ and $\alpha$. Above a certain critical value of $\alpha$ we observe the periodic pattern characteristic of Faraday waves (FW), which arises following a time of the order of 20~ms after the excitation is applied. It manifests as a periodic pattern along the axial direction of the cloud, as shown in Fig.~\ref{fig:fig1}(a). This pattern has a specific wavevector $k_{FW}$, as revealed by its spatial Fourier transform shown in Fig.~\ref{fig:fig1}(b). Similar images are obtained for all the explored values of $1/k_Fa_s$. The Faraday patterns exhibit best visibility when the trap is cylindrically symmetric because the excitation process is resonant with both radial frequencies. For this reason, we employ time-averaged potentials \cite{Roy} to produce a perfectly cylindrical trap.

In order to perform a systematic analysis, we excite the sample and observe its dynamics during an interval of 60~ms, longer than the other time-scales of the system (i.e. the period of the trap $2\pi/\omega_r^0 \sim 6$~ms and the excitation time $\tau_e \simeq 30.7$~ms). We observe these dynamics for different values of the excitation amplitude $\alpha$ and for different values of the interaction parameter $1/k_Fa_s$. To precisely track the dynamics of the excited cloud, for a given value of the interaction parameter $1/k_Fa_s$, we acquire images with a time step of $\Delta t = 0.5$~ms and for every value of the evolution time $t$, we take the average of five different images under same conditions.

\begin{figure}
\centering
 \includegraphics[width=\columnwidth]{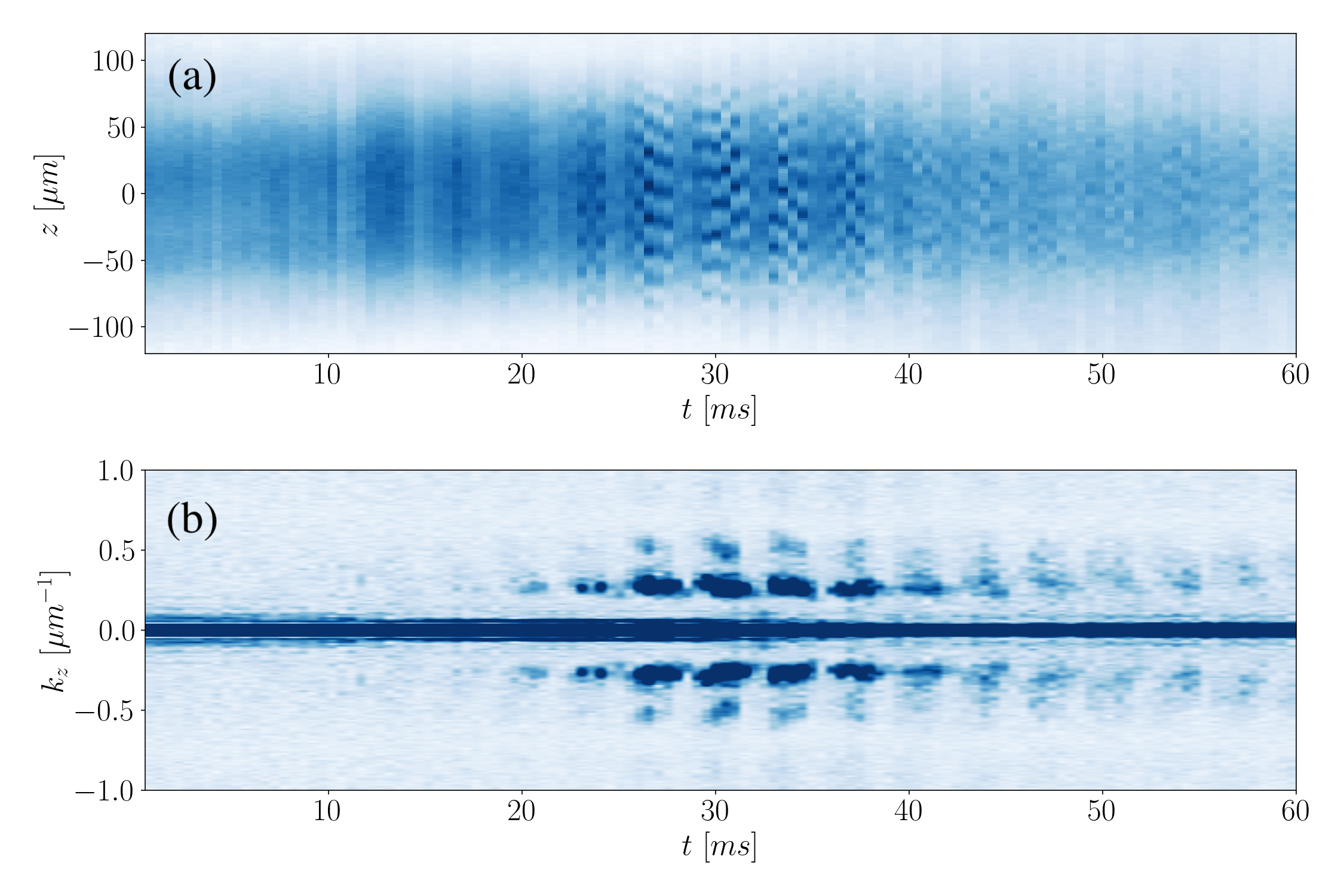} 
 \includegraphics[width=\columnwidth]{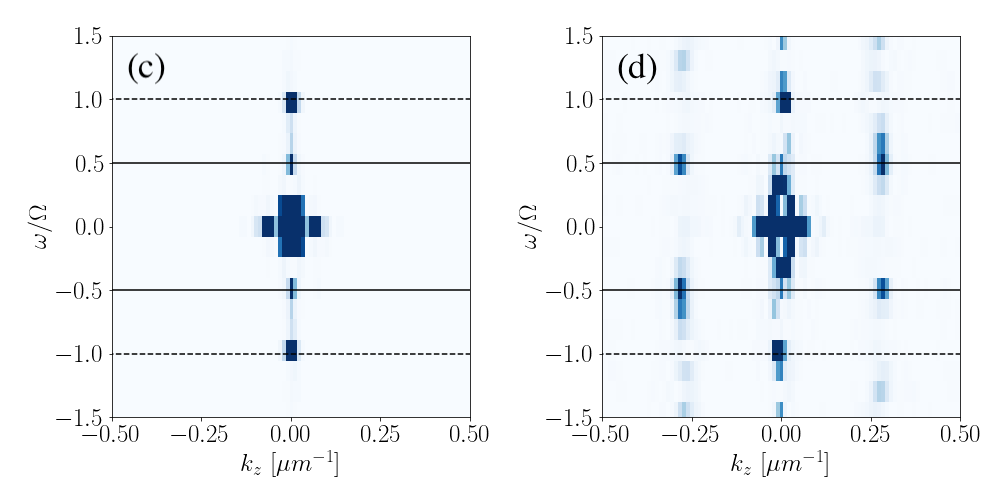} 
 \caption{Time evolution of the measured (a) integrated optical density and (b) its spatial Fourier transform. Panels (c) and (d) show the spatial-temporal Fourier transform of (a) before ($t<20$~ms) and after the onset of the Faraday pattern ($20<t<40$~ms), respectively. The dashed lines indicate the position of the breathing mode frequency, while the solid lines correspond to the Faraday temporal frequency. All measurements are taken for $1/k_Fa_s = 7.1$ (690~G) and $\alpha = 0.17$.}
\label{fig:fig2}
\end{figure}

To easily visualize all the data, we integrate along the radial direction the optical density and its Fourier transform, and create the time series shown in Fig.~\ref{fig:fig2}(a) and (b), where we can see the axial profile as a function of time. In these series, the Faraday patterns can be seen as a periodic structure. Fig.~\ref{fig:fig2}(b) shows the spatial Fourier transform of this series where we can clearly observe the appearance of the Faraday pattern as a separate $k$-component. It can also be observed that once the FW emerge, they undergo a death-revival sequence with a frequency  equal to $\Omega/2$, which is characteristic of this phenomenon.

The space-time Fourier transform of Fig.~\ref{fig:fig2}(a) is shown in Figures~\ref{fig:fig2}(c) and (d), providing insightful information of the dynamics of the system. Fig.~\ref{fig:fig2}(c) shows the space-time FFT of the system before the onset of the Faraday waves ($t<20$~ms). Two peaks can be observed, each one at $k_z = 0$ and $\omega = \pm \Omega$ which correspond to the breathing/driving mode (indicated by the dashed lines). However, for a later time, the emergence of the Faraday patterns can be seen in Fig.~\ref{fig:fig2}(d) by the appearance of four new peaks at approximately $k_z \approx \pm 0.25~\mu m^ {-1}$ and $\omega = \Omega/2$ (indicated by the solid lines). This is a clear signal of the onset of the Faraday instability.

\begin{figure}
\centering
 \includegraphics[width=\columnwidth]{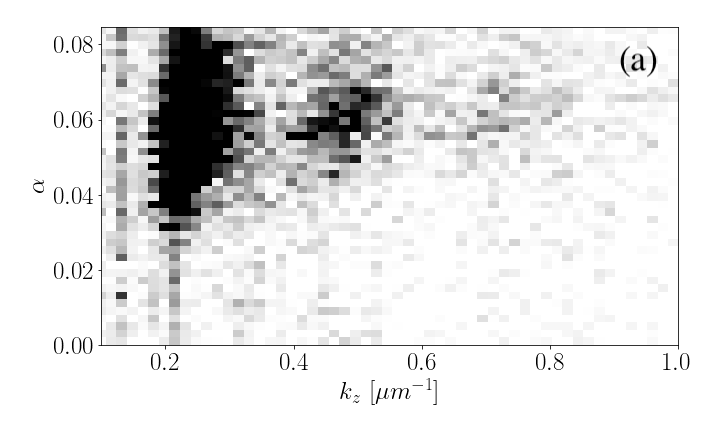}
 \includegraphics[width=\columnwidth]{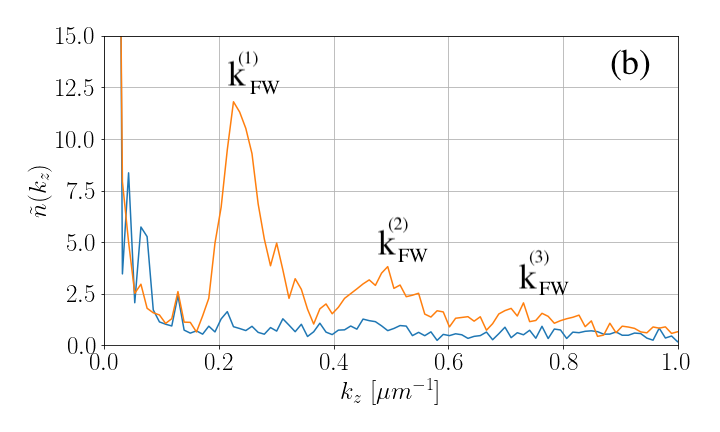}
 \caption{(a) Radially integrated Fourier transform for different excitation amplitudes. (b) Intensity of the $k$-components of a supefluid with Faraday waves for an an excitation amplitude $\alpha = 0.06$ (orange curve) and its comparison with a superfluid with no excitation (blue curve). Three peaks, $k^{(1)}_{FW}$, $k^{(2)}_{FW}$, and $k^{(3)}_{FW}$, can be clearly observed. Here $1/k_Fa_s = 7.1$ (690~G), similar images are obtained for other values of the interaction parameter.}
\label{fig:fig3}
\end{figure}

By increasing the value of the excitation amplitude $\alpha$, we observe two things: (i) the Faraday pattern appears at a shorter time and (ii) higher $k$-components appear. We are able to clearly observe a second $k$-component (see Fig.~\ref{fig:fig2}(b)) and, in some cases, even a third one. For this reason, we denote the Faraday wavevector as $k^{(n)}_{FW}$, where the superscript $^{(n)}$ indicates the $n$-th $k$-peak. To characterize the effect of $\alpha$, we select a time in which the Faraday pattern is fully formed in the Fourier space (for instance, at $t=31.4$~ms in Figure~\ref{fig:fig2}(b)) and plot that profile for different values of $\alpha$, generating a $k - \alpha$ diagram shown in Figure~\ref{fig:fig3}(a) where three different $k$-components ($k^{(1)}_{FW}$, $k^{(2)}_{FW}$, and $k^{(3)}_{FW}$) are clearly distinguished. Similar diagrams are obtained for other values of $1/k_Fa_s$, however, the contrast and the number of Faraday components decrease as the interaction strength increases (see further discussion in Sec.~\ref{sec:strong}). 

In the diagram of Fig.~\ref{fig:fig3}(a) the peak corresponding to $k^{(1)}_{FW}$ appears broader than the other two peaks, this is an effect associated to the difference between their relative heights. To clarify this point we present Fig.~\ref{fig:fig3}(b) where we plot the intensity of the Faraday peaks (orange curve) for $\alpha = 0.06$. Hence, in order to make visible all three $k$-components in Fig.~\ref{fig:fig3}(a), we need to include $k_{FW}^{(1)}$ from the base, where it is broader. For comparison, the blue curve in Fig.~\ref{fig:fig3}(b) corresponds to a superfluid with no excitations.

This diagram is very important since it can be interpreted as a stability diagram showing the conditions under which Faraday waves can be formed. Indeed, this is reminiscent of the stability diagram shown in reference \cite{Staliunas2002} for a 2D BEC.

To support our data, we perform numerical simulations by solving the time-dependent Gross-Pitaevskii equation (GPE) \cite{Pethick} in which the radial frequency is modulated as explained above, that is $\omega_r(t) = \omega_r^0 \sqrt{1 + \alpha\sin(\Omega t) }$. As pointed out in references \cite{Staliunas2002, Staliunas2004, Okazaki}, dissipation is an important condition to generate the patterns. To account for dissipation, we introduce a dimensionless parameter $\gamma_d$ that can be interpreted as an imaginary time parameter \cite{Choi, Kagan}. Under these considerations, the GPE we use is:

\begin{eqnarray}	\label{eq:GPE}
(i - \gamma_d) \hbar\frac{\partial \Psi}{\partial t} = \left[-\frac{\hbar^2}{2M}\nabla^2 + \frac{M}{2}\left( \omega_r^2(t)r^2 + \omega_z^2z^2 \right) + \right. \nonumber \\
   \left. + \frac{4\pi \hbar^2 a_M}{M} \left| \Psi \right|^2  \right] \Psi 
\end{eqnarray}
where $M$ is the mass of an atomic pair ($M = 2m$) and $a_M$ is the inter-molecular scattering length, $a_M \simeq 0.6a_s$ \cite{Petrov}. For $\gamma_d = 0$ FW still occur, but this requires a longer evolution time and the system becomes unstable due to the parametric nature of the excitation. To reduce the evolution time, we introduce a small numerical noise in the initial state and set $\gamma_d = 0.01$ to stabilize the system.

\begin{figure}
\centering
 \includegraphics[width=\columnwidth]{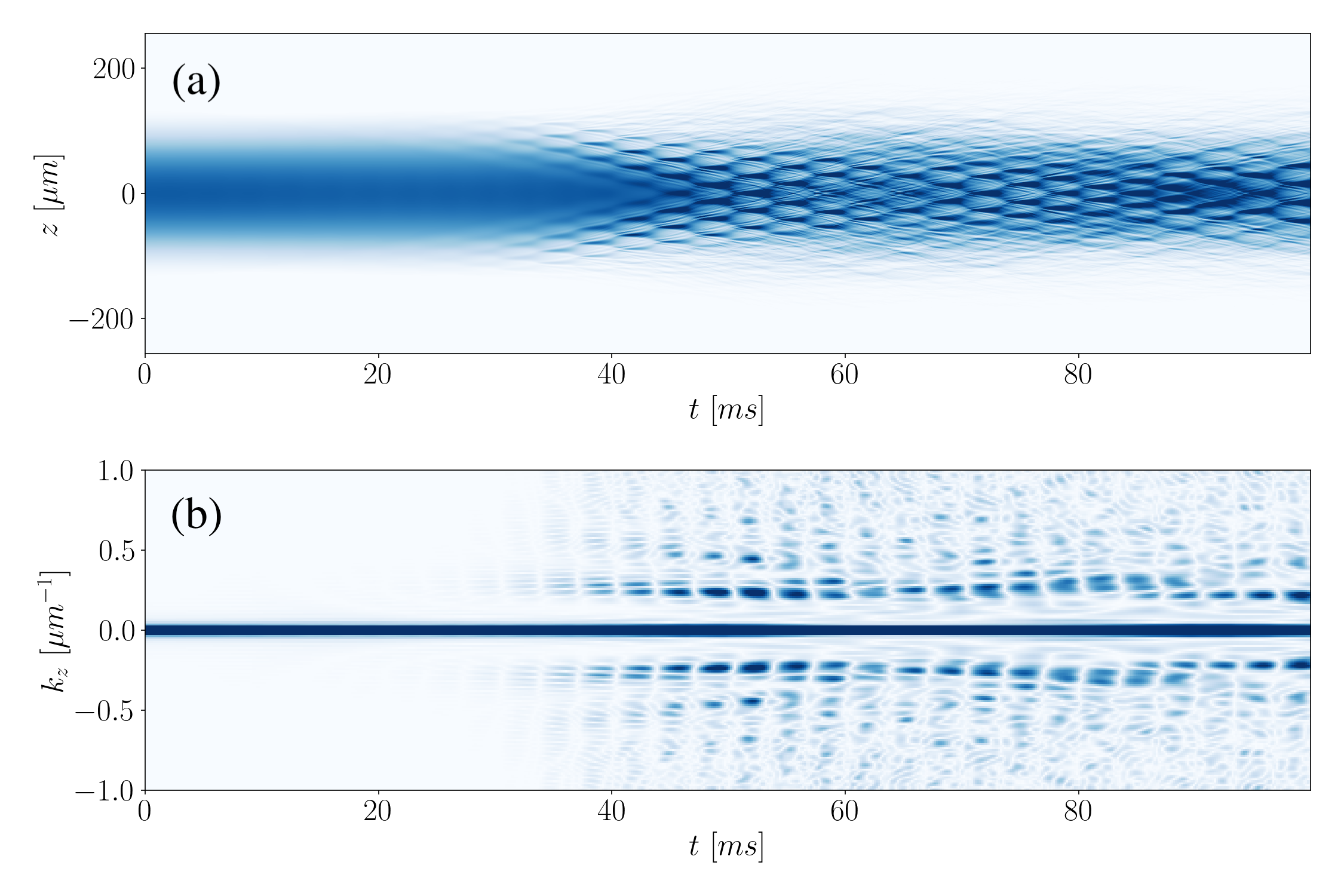} 
 \includegraphics[width=\columnwidth]{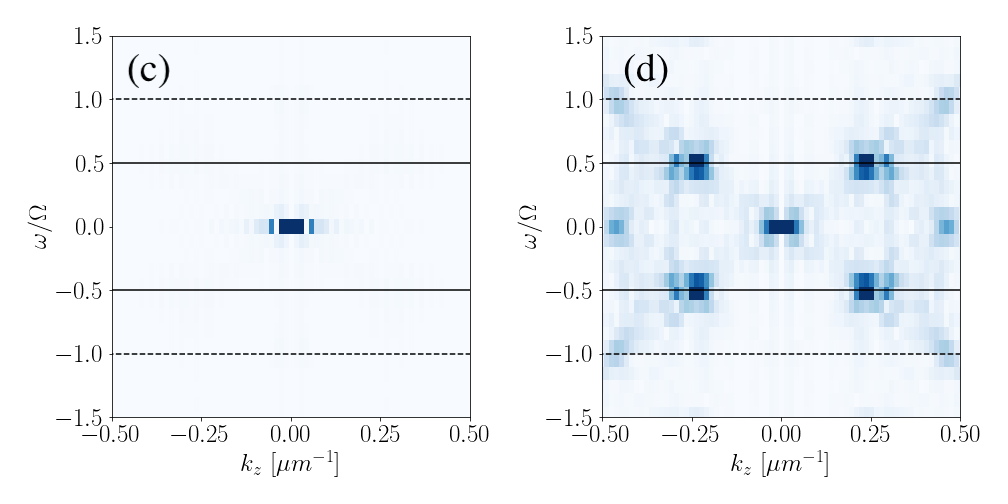} 
 \caption{Numerical simulation of the time dependent Gross-Pitaevskii equation. Panel (a) shows the integrated density of the superfluid and (b) its spatial Fourier transform as a function of time. Panels (c) and (d) show the spatial-temporal Fourier transform of (a) before ($t<25$~ms) and after ($25\mbox{ ms}<t<43$~ms) the onset of the Faraday pattern, respectively. The dashed lines indicate the position of the breathing mode frequency, while the solid lines correspond to the Faraday waves temporal frequency. In these simulations $1/k_Fa_s = 7.1$, $\gamma_d = 0.01$ and $\alpha = 0.17$.}
\label{fig:fig4}
\end{figure}

To solve Eq.~(\ref{eq:GPE}) we use a spectral split-step method to evolve the equation as a function of time and solve it under similar conditions as in the experiments (see for instance \cite{Bao}). We found that for $\gamma_d = 0.01$ our simulations correctly reproduce our observations. Figures~\ref{fig:fig4}(a) and (b) show the evolution of the radially integrated sample, as well as its Fourier transform, respectively. Figs.~\ref{fig:fig4}(c) and (d) show the spatial-temporal Fourier transform before ($t<25$~ms) and after ($25\mbox{ ms}<t<43$~ms) the onset of the Faraday pattern, respectively. Comparing this results with Figure~\ref{fig:fig2} we can see that the agreement is remarkable.

It is important to mention that although our simulations also show the non-linear regime reported in references \cite{Nguyen,Okazaki} for longer evolution times or larger excitation amplitudes, in the experiment we observe a  destruction of the superfluid state as a clear reduction of the condensed fraction. This result is to be expected due to the parametric heating nature of the excitation. Moreover, to explore the non-linear regime observed in \cite{Okazaki} it would be best to acquire a momentum resolved image rather than a resolution limited in situ image. We are indeed interested in developing such imaging techniques an may be the an interesting topic for future research. Additionally, we do not observe the granulation state observed in \cite{Nguyen} for which a key ingredient is the use of lower frequencies, $\Omega<\omega_r$, and large excitation amplitudes.

\begin{figure}
\centering
	\includegraphics[width=\columnwidth]{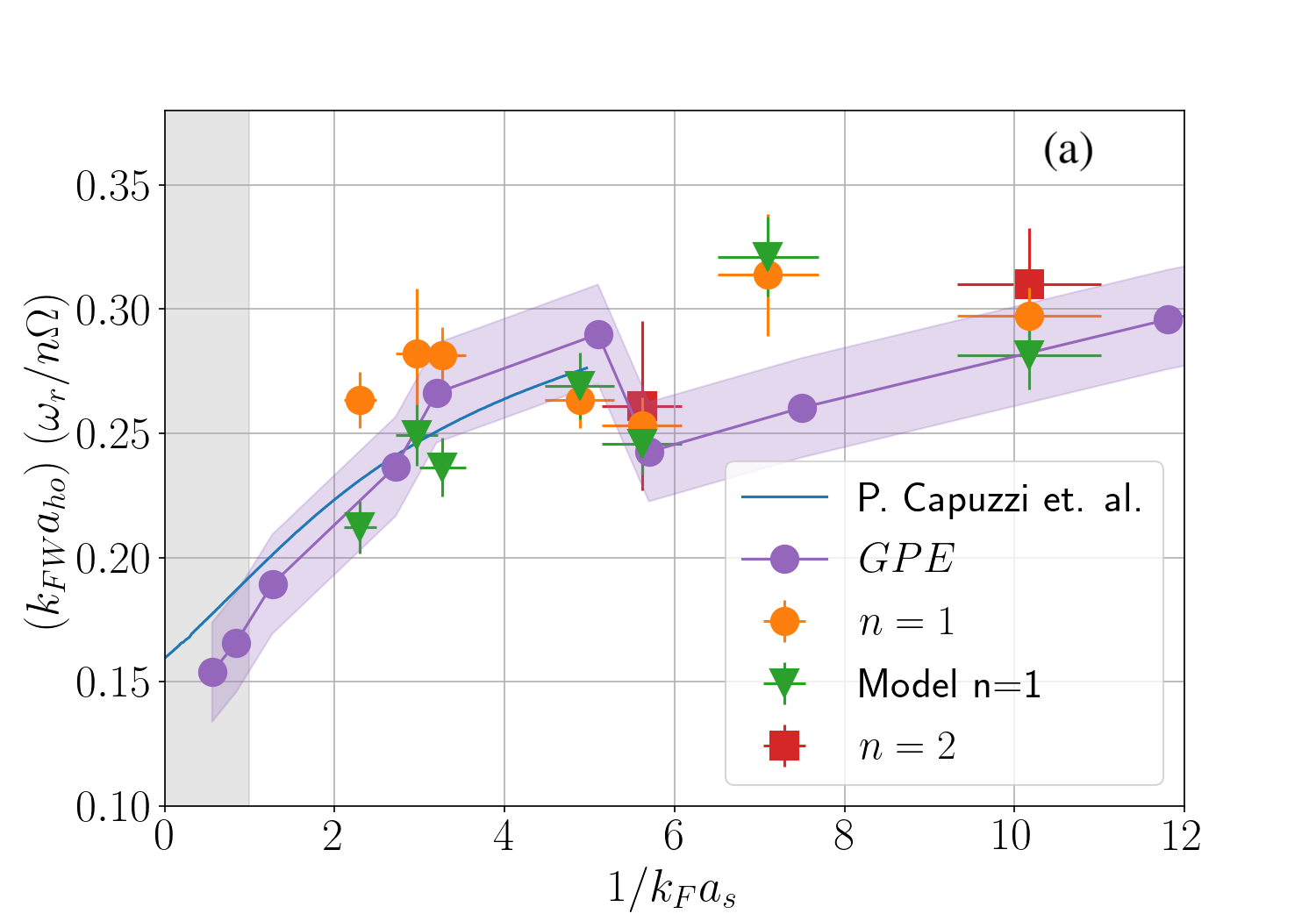}
	\includegraphics[width=\columnwidth]{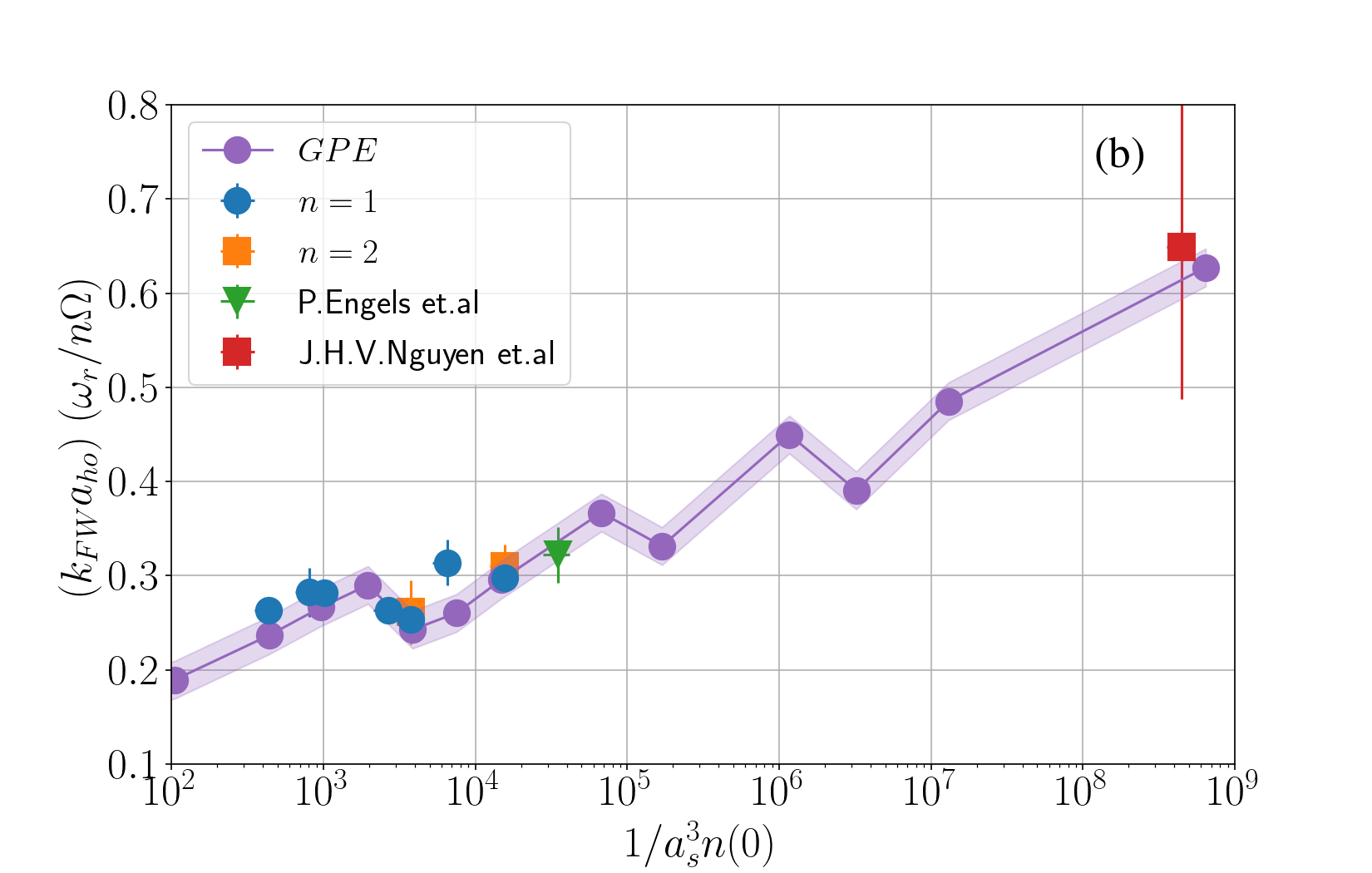}
 \caption{(a) Measurement of the Faraday wavevector. Here we plot the product $k^{(n)}_{FW} a_{ho} \omega_r / (n\Omega)$ (orange circles and red squares) so it can be compared with the model from \cite{Capuzzi} (solid blue curve). The green triangles correspond to the Floquet model from Section~\ref{sec:floquet}. (b) Comparison of our data with other parametrically excited BEC systems composed by $^7$Li \cite{Nguyen} (red square), and $^{87}$Rb \cite{Engels} (green inverted triangle), as a function of the interacting parameter $a_s^3n(0)$. In both panels the purple dots correspond to the results from the GPE simulation while the shaded region is associated to the width of the first Faraday peak. The error bars of our data correspond to the width of the $n$-th Faraday peak.} 
\label{fig:fig5}
\end{figure}

From these measurements and simulations, we can determine the wave vector $k^{(n)}_{FW}$ associated to the Faraday pattern as a function of the interaction strength. Inspired by Fig. 7 from reference \cite{Capuzzi}, in Fig.~\ref{fig:fig5}(a) we plot the product, $k^{(n)}_{FW} a_{ho} \omega_r/(n \Omega)$, where $a_{ho} = \sqrt{\hbar / m\omega_r}$ is the harmonic oscillator length along the radial direction. This quantity decreases as the interaction strength increases, meaning that the pattern spacing increases with the scattering length. This is consistent with the $k^{(n)}_{FW}$ extracted from our GPE simulations (purple points, the purple shaded region corresponds to the error associated to the width of the first Faraday peak), as well as with the Floquet analysis (green triangles) from section~\ref{sec:floquet}. The solid blue curve in Fig.~\ref{fig:fig5}(a) corresponds to the solution of the superfluid hydrodynamic equations presented in reference \cite{Capuzzi}, showing good agreement with the experiment for values of $1/k_F a_s < 5$.

In order to compare our observations with previous experiments where FW were parametrically excited, namely $^7$Li \cite{Nguyen} and $^{87}$Rb \cite{Engels}, we normalized the interaction parameter as $1/a_s^3n(0)$ where $n(0)$ is the peak density of the BEC at the center of the trap \cite{Dalfovo} and compare them with our data ($k^{(1)}_{FW}$, blue circles, and $k^{(2)}_{FW}/n$, orange squares) in Fig.~\ref{fig:fig5}(b). We have, in fact, accessed a regime two orders of magnitude more interacting than what has been previously studied. One can also observe that the dimensionless parameter $k^{(n)}_{FW} a_{ho} \omega_r/(n \Omega)$ decreases logarithmically as the interaction parameter increases.

We can now use the obtained Faraday wavevector to extract the associated phase velocity of the Faraday waves as a function of the interaction parameter. A relevant quantity is the phase velocity of the $n$-th excitation given by
\begin{equation}\label{eq:phase}
v^{(n)}_{ph}=\frac{\omega_{FW}}{k_{FW}} = \frac{n\Omega/2}{k_{FW}}
\end{equation}

\begin{figure}
\centering
 \includegraphics[width=\columnwidth]{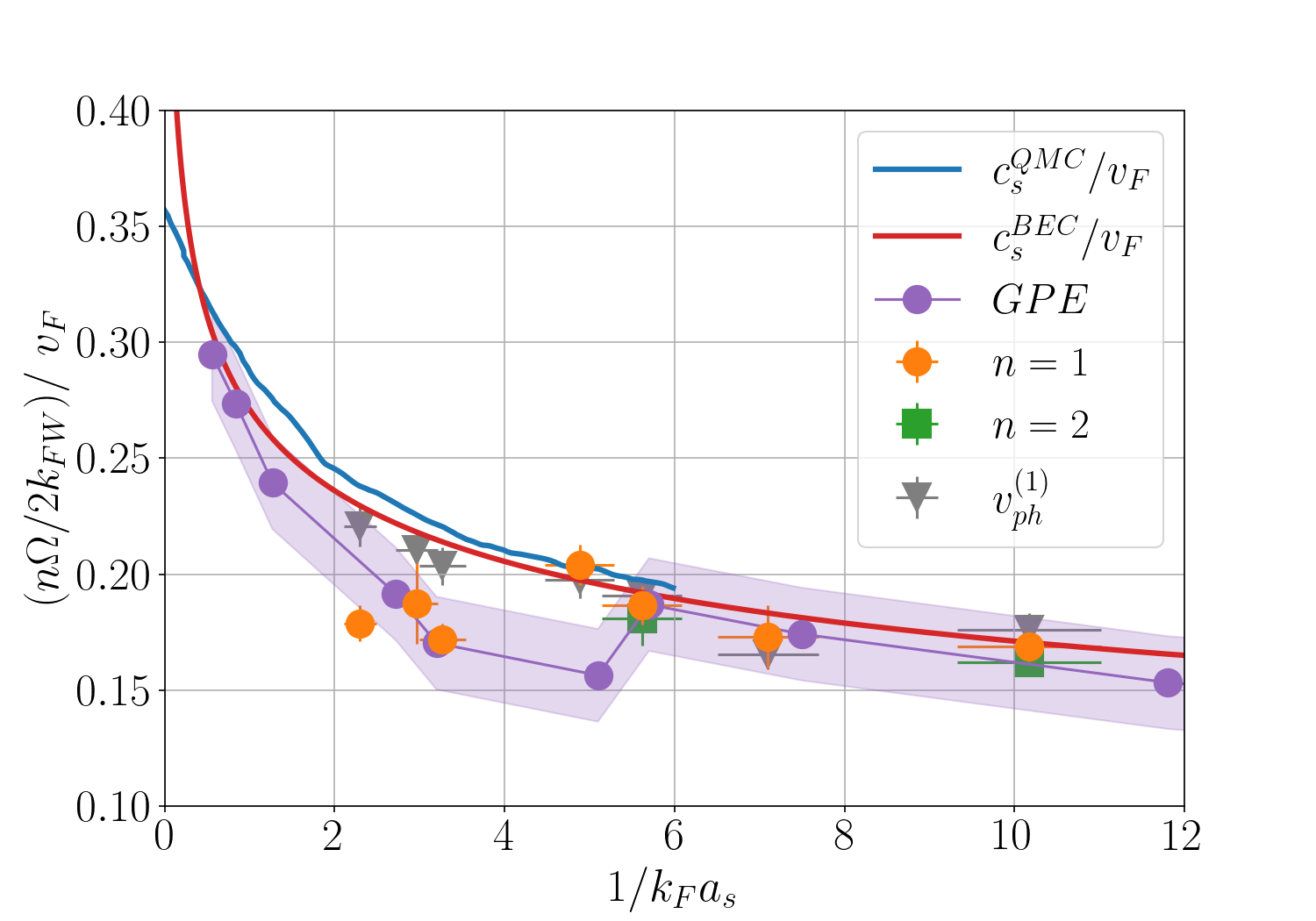}
 \caption{Measured phase velocity of FW as a function of the interaction parameter (orange circles and green squares). Different models (see text) indicate that this velocity corresponds to the effective 1D superfluid speed of sound. The error bars of our data correspond to the width of the $n$-th Faraday peak.} 
\label{fig:fig6}
\end{figure}

We claim that this phase velocity corresponds to the effective 1D speed of sound of the superfluid. To support this claim, we plot our data in Fig.~\ref{fig:fig6} (orange circles and green squares) and compare them with the expected speed of sound according to different models (we scale them with the Fermi velocity, $v_F$). The red curve represents the speed of sound of a Bose-Einstein condensate, that is $c_s^{BEC} = \sqrt{gn(0)/M}$, where $n(0)$ is the peak density of the molecular gas at the center of the trap \cite{Pethick}. We also compare our data with a more general model (blue curve) corresponding to the experimentally agreeing results from Quantum Monte Carlo calculations \cite{Giorgini, Joseph}. The results from the GPE simulations are shown by the purple points and shaded region. Finally, we also compare our data with a simple model based on Floquet theory (blue diamonds) which we will discuss with more detail in Section~\ref{sec:floquet}.

We see that the agreement between our data and theory is very good considering that no fitting parameters were employed in Fig.~\ref{fig:fig6}. However, in general our data points lays below the expected speed of sound, a more detailed analysis of the physics involving this process needs to be carried out. As pointed out by M. Mossman et al. \cite{Mossman}, highly excited superfluid gases may exhibit an effective lower speed of sound.

This result is important since it provides a new and simple way to measure the speed of sound of the superfluid, complementing other available methods \cite{Sidorenkov, Joseph, Biss}. 

\section{Unitary regime}\label{sec:strong}

Being a strongly correlated regime, observing FW at unitarity is certainly very interesting. We applied the same protocol that we employed on the molecular condensate to generate FW at or near unitarity. Although we do excite a breathing mode, we could not observe a Faraday pattern in the absorption images. 

The numerical simulations from references \cite{Capuzzi, Tang} indicate that Faraday waves should be formed throughout the crossover. Moreover, we have performed numerical simulations using the extended Thomas-Fermi model which describes a Fermi superfluid at unitarity where the scattering length diverges \cite{Manini,Salasnich,Forbes}. It consists of a mean-field, zero temperature model in which the order parameter $\Psi$ satisfies the following equation:

\begin{eqnarray}	\label{eq:ETF}
(i - \gamma_d) \hbar\frac{\partial \Psi}{\partial t} = \left[ -\frac{\hbar^2}{2M}\nabla^2 + \frac{M}{2}\left( \omega_r^2(t)r^2 + \omega_z^2z^2 \right) + \right. \nonumber \\
   \left. + \xi\frac{\hbar^2}{M}(3\pi^2)^{2/3} \left| \Psi \right|^{4/3}  \right] \Psi \ \
\end{eqnarray}
where $M$ is the mass of an atomic pair, $\xi \simeq 0.370$ is the Bertsch parameter \cite{Ku, Jauregui} and $\gamma_d$ is the dissipation parameter.

\begin{figure}[!ht]
\centering
 \includegraphics[width=\columnwidth]{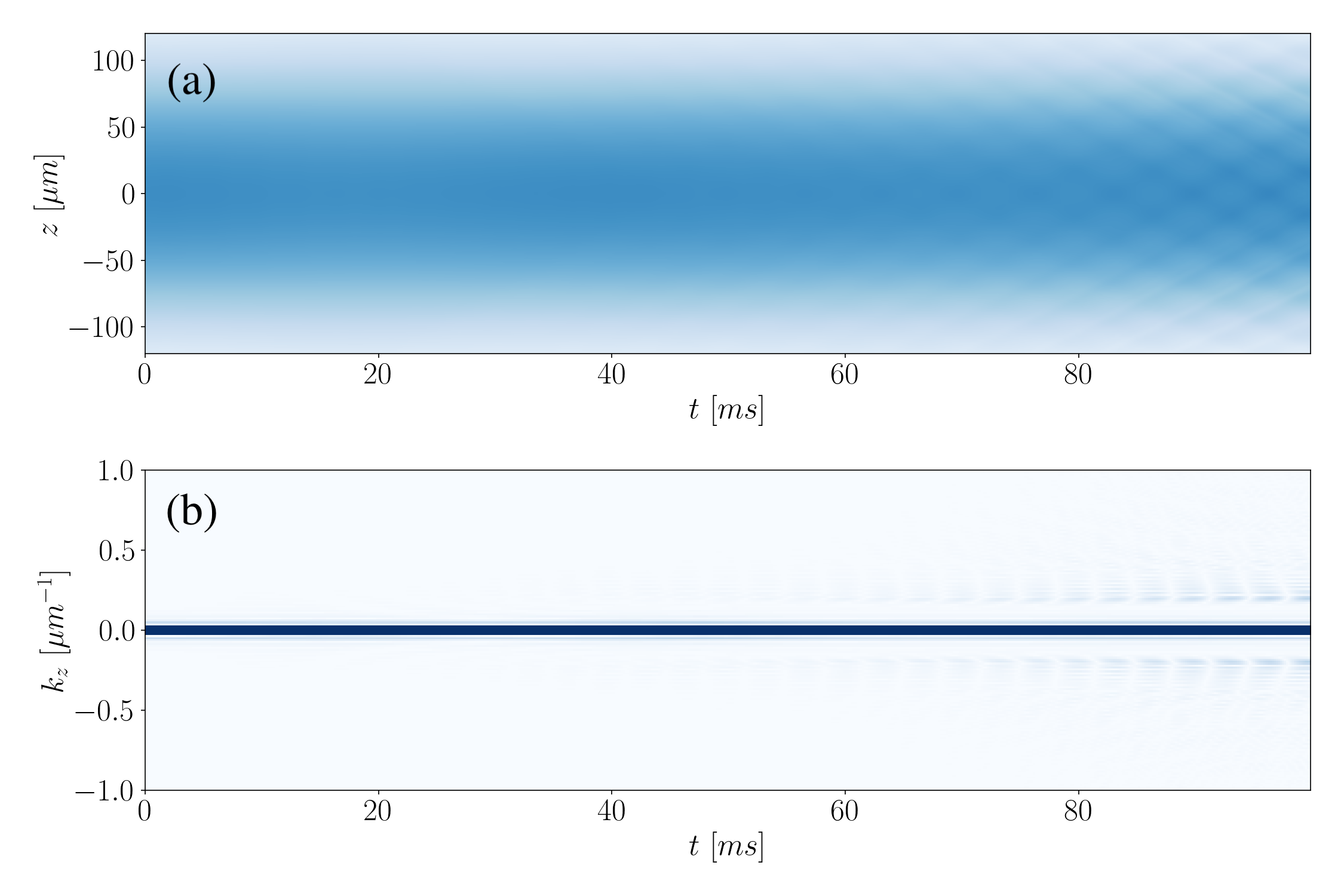}
 \includegraphics[width=\columnwidth]{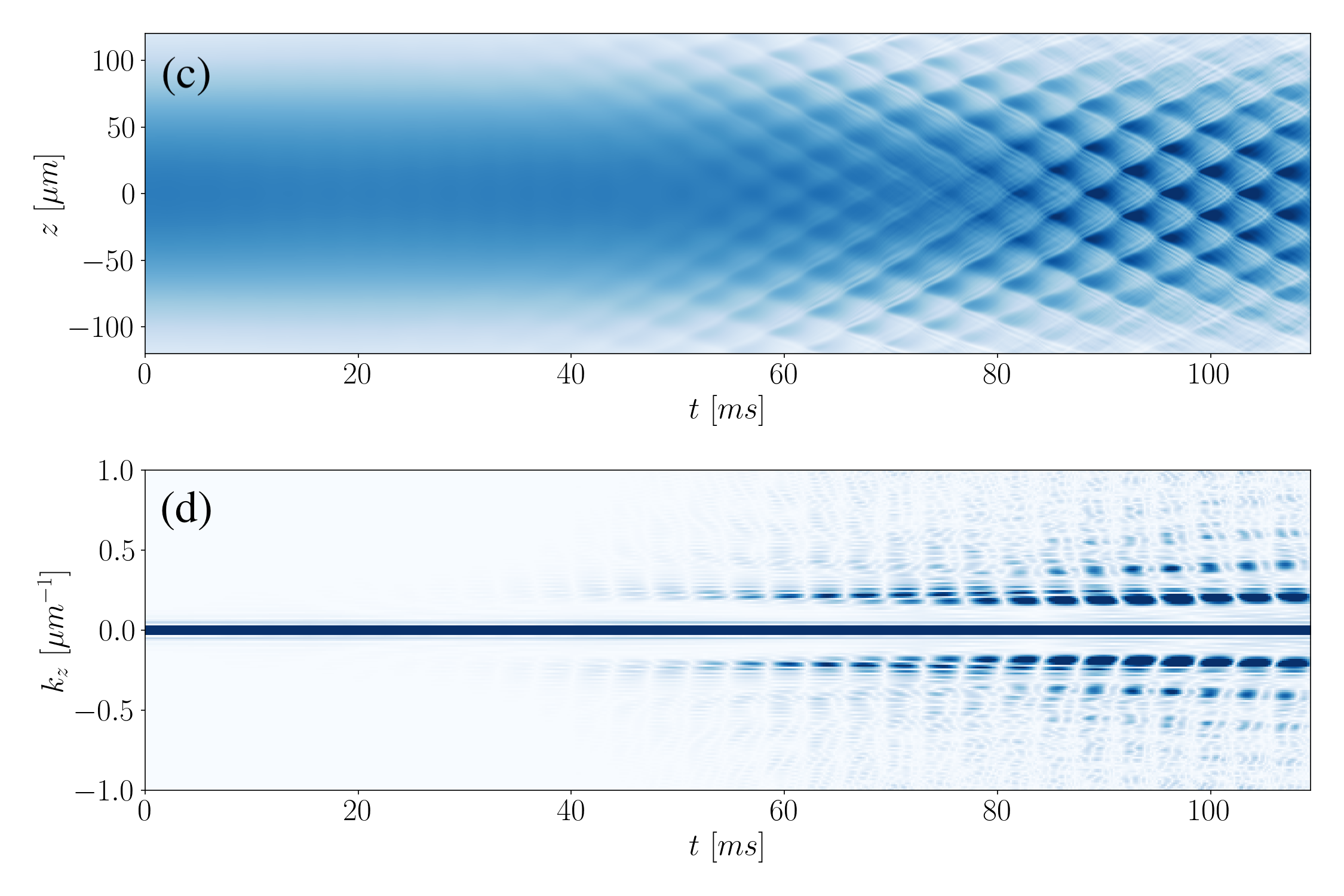}
 \includegraphics[width=\columnwidth]{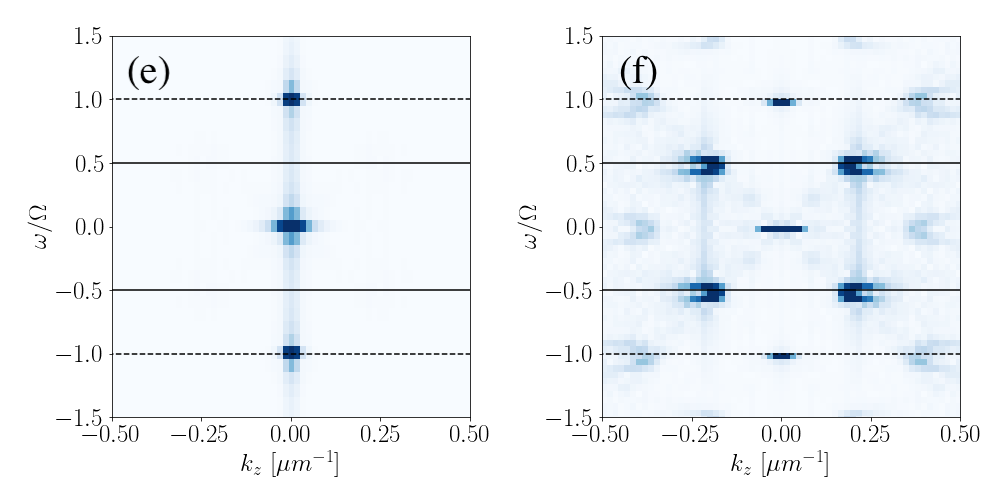}
 \caption{Numerical simulations of the extended Thomas-Fermi model at unitarity. Panels (a) and (b) show, respectively, the time evolution of the integrated density and spatial Fourier transform for an excitation frequency of $\Omega = 2\omega_r^0$. Panels (c) and (d) show the same quantities for an excitation frequency that matches the breathing mode at this regime, $\Omega = \sqrt{10/3}\,\omega_r^0$. Panels (e) and (f) show the spatial-temporal Fourier transform of (c) before ($t<40$~ms) and after ($40\mbox{ ms}<t<60$~ms) the onset of the Faraday pattern, respectively. Dashed lines indicate the position of the breathing mode frequency, while the solid lines correspond to the Faraday waves temporal frequency. In these simulations $1/k_Fa_s = 0$, $\gamma_d = 0.01$ and $\alpha = 0.1$.} 
\label{fig:fig7}
\end{figure}

Using the same value of the modulation frequency as before, $\Omega = 2\omega_r^0$, generates a barely visible Faraday pattern in the experimentally accessible timescales before observing the destruction of the sample. This faint patterns are shown in Figs.~\ref{fig:fig7}(a) and (b). However, when the excitation frequency matches the breathing mode frequency of the unitary regime, $\Omega = \omega_b = \sqrt{10/3}\,\omega_r^0$ \cite{Giorgini, Heiselberg, Manini}, the Faraday pattern emerges very clearly, as can be seen in Fig.~\ref{fig:fig7}(c), (d) and (f). Indeed, numerical simulations suggest that FW at unitarity are very sensitive to the value of the excitation frequency. This is not a surprise since the unitary gas is much more incompressible and, therefore, more rigid than a BEC \cite{Ku}, making the system response narrower and FW harder to observe.

According to these mean field models \cite{Capuzzi,Tang}, it should be possible to excite and observe FW across the BEC-BCS crossover. Then, the question that arises is why don't we observe them in the experiment. It is important to consider the fact that these models assume that the condensed fraction is 100\% at unitarity. However, this is not the case in a real unitary Fermi gas, where we know that even at $T=0$ the maximum condensed fraction that can be obtained is 50\% due to beyond mean-field effects associated to strong interactions \cite{Giorgini, Ku, Inada}. 

Hence, one possibility is that Faraday excitations are being produced but they are not visible due to the presence of a very large non-condensed fraction that hides the pattern, just as it happens when quantized vortices are nucleated at strong interacting regimes \cite{Zwierlein}. To support this hypothesis we studied the contrast of the Faraday pattern as a function of temperature for different values of the interacting parameter. 

\begin{figure}
\centering
 \includegraphics[width=\columnwidth]{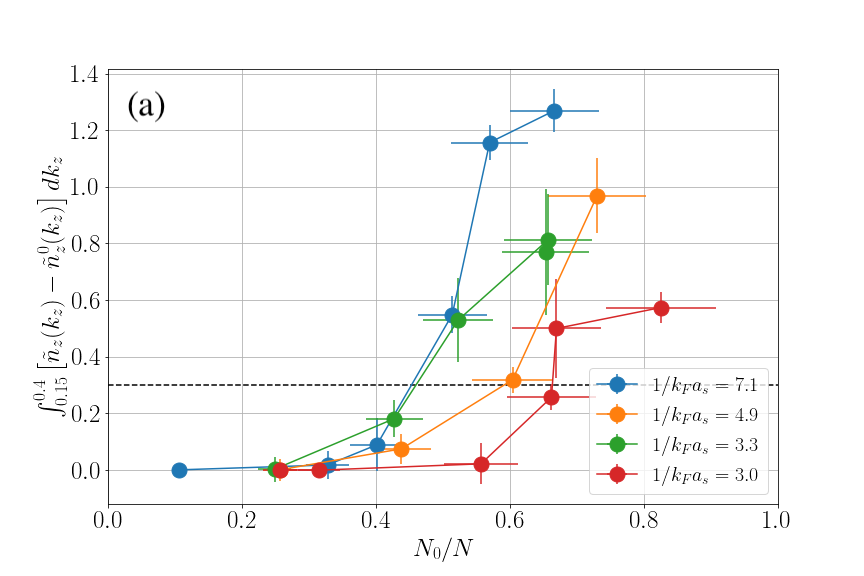}
 \includegraphics[width=\columnwidth]{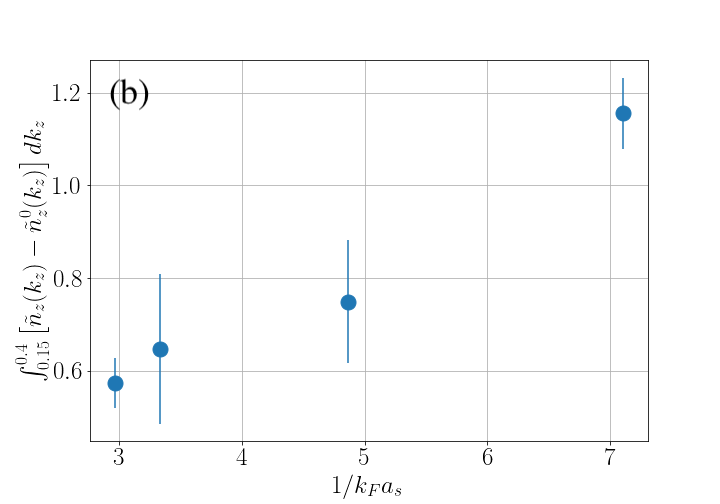}
 \caption{(a) Integrated Fourier signal of the $k_{FW}^{(1)}$ peak as a function of the condensed fraction for different values of the interacting parameter $1/k_Fa_s$. Below the dashed line the pattern is not visible anymore. (b) Maximum observed value of $\mathcal{C}$ at lowest temperature as a function of $1/k_Fa_s$. Error bars correspond to the standard deviation of five independent measurements.} 
\label{fig:fig8}
\end{figure}

We estimate this contrast by calculating the integrated Fourier signal associated to the first Faraday peak $k_{FW}^{(1)}$ using the formula

\begin{equation}\label{eq:contrast}
\mathcal{C} = \int \left( \tilde{n}_z(k_z) - \tilde{n}_z^0(k_z) \right) dk_z,
\end{equation}
where $\tilde{n}_z^0(k_z)$ and $\tilde{n}_z(k_z)$ are the radially integrated spatial Fourier transform of the density profile before and  after the Faraday pattern is formed, respectively. In Eq.~\ref{eq:contrast}, the integration is performed in the region around the first Faraday peak. The results are shown in Figure~\ref{fig:fig8}(a), where we can see how the contrast of the pattern rapidly drops as the condensed fraction decreases. Below the dashed line the Faraday pattern is not visible anymore, however $\mathcal{C}$ is not zero due to the reminiscent effect of the change in the cloud's width in the Fourier transform. Moreover, in Figure~\ref{fig:fig8}(b) we can see that the maximum observed contrast (at lowest temperature) decreases as the system becomes more interacting. This supports the idea that FW are not observed at unitarity due to the presence of a large non-condensed cloud.

From the numerical simulations of the GPE-like equation~(\ref{eq:ETF}), we observe that even exciting at resonance with the breathing mode of the unitary superfluid, the growth rate of the FW excitation is lower than in the BEC, comparing Figs. \ref{fig:fig4}(b) and \ref{fig:fig7}(c), as already noted by R.-A. Tang et al. \cite{Tang}. This result suggest that the excitation duration should be longer. Moreover, being a parametric process, the excitation will tend to heat up the sample \cite{Savard,Friebel}. This effect is stronger in a less compressible system, where this kind of excitations are harder to produce and hence the pumped energy will be more easily employed to heat up the gas, which will result in a decrease in the condensed fraction. Indeed, we have measured the temperature of the sample after the excitation is applied, we do observe an important increase of the temperature after 20 cycles of excitation. Therefore, it is also feasible that we are not able to produce the Faraday patterns at unitarity at all, and an alternative protocol may be needed.

This indicates that our protocol might not be the best to generate this excitations at unitarity, and a different scheme is required. Perhaps, a quench in the trapping potential rather than a parametric process, like the one described in \cite{Smits}, could generate the pattern without significantly increasing the temperature of the system. To unambiguously detect the Faraday pattern at unitarity we will require momentum resolving techniques, such as Bragg spectroscopy \cite{Ernst, Fabbri}. We expect that this technique would show a signal located at the corresponding Faraday wavevector $k_{FW}$, indicating the presence of the excitation. This remains to be investigated in a future study.

\section{Floquet analysis and stability diagram}\label{sec:floquet}

From the discussion in Section~\ref{sec:fw}, we see that the Faraday patterns on the BEC side of the resonance are well described by the Gross-Pitaevskii equation. In this section, we present a different analysis which allows us to investigate the phenomenon in terms of stability.

In classical systems, Faraday waves can be explained as unstable solutions of the Mathieu equation \cite{Cross}. This is also the case in our superfluid. To see this, we introduce an analytic Floquet model similar to the one presented in reference \cite{Hai}. We start by imposing a condition in the GPE~(\ref{eq:GPE}) in which the potential and the mean-field interaction energies are balanced, that is,
\begin{equation}\label{eq:balance}
V(\vec{r}, t) + g\left| \Psi(\vec{r}, t) \right|^2 = E_F,
\end{equation}
where $g = 4\pi \hbar^2 a_M / M$ is the interaction parameter and the potential energy is given by
\begin{eqnarray}\label{eq:potential}
V(\vec{r}, t)  &=& \frac{M}{2}\left( \omega_r^2(t)r^2 + \omega_z^2z^2 \right), \nonumber \\
\omega_r(t) &=& \omega_r^0 \sqrt{1 + \alpha\sin(\Omega t) }.
\end{eqnarray}

In the balance condition~(\ref{eq:balance}), $E_F$ is a constant that we will refer as the Floquet energy. This condition can be interpreted as the driving being an adiabatic-like process in which the variation of the potential is compensated by the density change during the whole dynamics of the system. Notice that this condition implies that at any time the density at $r=0$ is kept constant. This is not the case in the experiment, nonetheless, this simplified model gives a qualitative description of the process.

From Eqs.~(\ref{eq:potential}) and (\ref{eq:psiF}) we can see that $\Psi_F$ corresponds to the breathing/driving mode of the system. Using this balance condition, we find solutions to be
\begin{equation}\label{eq:psiF}
\Psi_F = e^{iS(\vec{r}, t)}\left( \frac{E_F - V(\vec{r}, t)}{g} \right)^{1/2},
\end{equation}
where the phase $S(\vec{r}, t)$ is chosen such that the equation
\begin{equation}\label{eq:GPE_F}
i \hbar\frac{\partial \Psi_F}{\partial t} = -\frac{\hbar^2}{2M}\nabla^2\Psi_F + E_F\Psi_F 
\end{equation}
is satisfied. We assume that this equation has a non trivial solution. It is important to mention that the balance condition imposes a dissipationless equation in which $\gamma_d = 0$.

We investigate the stability of the Floquet sate~(\ref{eq:psiF}) using linear stability analysis. To do so, we introduce a small perturbation onto the Floquet states of the form $\Psi = \Psi_F\Psi_P$, where $\Psi_P = 1 + w(t)\cos(kz)$ is the perturbation. Here, $w(t)$ is a complex function given by $w(t) = u(t) + i v(t)$.

Now we substitute $\Psi$ into Eq.~(\ref{eq:GPE}), and consider the following approximations. First, we perform a linear stability analysis, meaning that we will neglect the second and higher order powers in $w(t)$. Next, we consider our trap to be harmonically confining along the radial direction, but to be homogeneous along the axial direction, i.e. $\partial \Psi_F / \partial z \simeq 0$. This is equivalent to consider the axial frequency of the trap to be zero, $\omega_z = 0$. Since we want to describe the Faraday pattern along the axial direction, only the time-varying part of the potential is important, so we spatially average the radial part, this is
\begin{equation}\label{eq:bulk}
\bar{V}(t) = \frac{M\omega^2_r R^2}{4} (1 + \alpha\sin\Omega t),
\end{equation}
where $R$ is the radial Thomas-Fermi radius. After this analysis we obtain the following equation for the real part of $w(t)$:

\begin{equation}\label{eq:interMathieu}
\ddot{u}(t) + \omega^2_k \left( \frac{E_k+2E_F}{E_k} - 2\frac{V(t)}{E_k} \right) u(t) = 0,
\end{equation}
where we have defined $E_k \equiv \hbar^2 k^2 /2M$ and $E_k \equiv \hbar\omega_k$. 

Notice that from the balance condition (\ref{eq:balance}), we can obtain the Floquet energy $E_F$ in different ways. By evaluating Eq.~(\ref{eq:balance}) at $z=0$, $r = R$ and, $t=0$ we obtain an expression for the Floquet energy: $E_F = M\omega^2_r R^2 / 2$. Simultaneously, by evaluating the balance condition at the center of the trap at $t = 0$ we obtain $E_F = g\left| \Psi_F (0) \right|^2$.

Finally, Eq.~(\ref{eq:interMathieu}) can be mapped into a Mathieu equation of the form
\begin{equation}\label{eq:Mathieu}
\frac{\partial^2 u (\tau)}{\partial \tau^2} + \left( a-2q\cos 2\tau \right) u(\tau) = 0,
\end{equation}
identifying,

\begin{eqnarray}
\tau &=& \frac{1}{2}\left(\Omega t - \frac{\pi}{2} \right), \label{eq:tau} \\ 
a &=& \frac{4}{\hbar^2\Omega^2}E_k(E_k + E_F), \label{eq:a} \\
q &=& \frac{4}{\Omega^2}\frac{E_kE_F}{2\hbar^2}\alpha. \label{eq:q}
\end{eqnarray}

The solutions of the Mathieu equation are well known for all values of $a$ and $q$, in fact, they can be categorized in stable and unstable solutions forming well defined regions in the $a-q$ diagram \cite{McLachlan}. Faraday waves can be identified as unstable solutions of the Mathieu equation.

We know that in the limit of small $q$ the instability ``tongues'' of the Mathieu equation are located at $a = n^2$ with $n$ an integer. Substituting this result in Eq.~(\ref{eq:a}), we find an expression for the Faraday wavevector $k^{(n)}_{FW}$,
\begin{equation}\label{eq:kFW}
k^{(n)}_{FW} = \pm \sqrt{\frac{ME_F}{\hbar^2}}\sqrt{-1+\sqrt{1 + \left( \frac{n\hbar\Omega}{E_F} \right)^2 }}.
\end{equation}

Here it is important to mention that although other approaches do arrive to the Mathieu equation (\ref{eq:Mathieu}) \cite{Staliunas2002, Staliunas2004, Nicolin2007, Nicolin2011} our approximation allows us to obtain an analytic expression for the Faraday wavevector (\ref{eq:kFW}) under the assumption of the balance condition of Eq.~(\ref{eq:balance}).

\begin{figure}
\centering
 \includegraphics[width=\columnwidth]{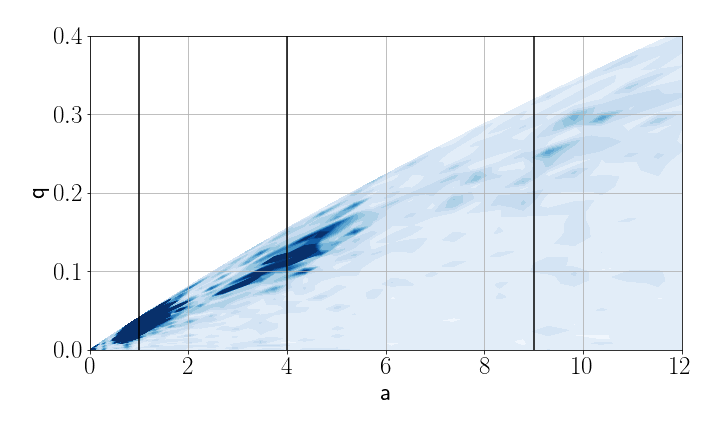}
 \caption{Stability diagram of the observed Faraday waves, the vertical lines correspond to the instability tingues of the Mathieu equation located at $a=n^2$ with $n=1,2,3$. Here $1/k_Fa_s = 7.1$ (690 G), similar images are obtained for other values of the interaction parameter. Here, the blank region is not experimentally accessible.} 
\label{fig:fig9}
\end{figure}

We now use expression~(\ref{eq:kFW}) and compare it with our experimental data and the GPE calculations shown in Fig.~\ref{fig:fig4}. The agreement is very good, indicating that our simple analytic model based on Floquet states provides a good qualitative description of this phenomenon. Nevertheless, it should be noted that this model slightly underestimates the value of $k^{(n)}_{FW}$ since it is consistently below the predictions from the hydrodynamic and GPE models, as can be seen in Fig.~\ref{fig:fig4}. This is due to the fact that the balance condition from Eq. (\ref{eq:balance}) considers that the density of the cloud at the center is unperturbed by the excitation.

Our model also provides an insightful interpretation of the parameters of Mathieu equation. We should notice that in the expression for $a$, the term $E_k(E_k + E_F)$ corresponds to the Bogoliubov phononic spectrum \cite{Pethick}, so Eq.~(\ref{eq:a}) can be rewritten as
\begin{equation} \label{eq:a2}
a = \frac{4}{\hbar^2\Omega^2}E_k(E_k + E_F) = \left( \frac{2\hbar \omega_{Bog}}{\hbar\Omega} \right)^2,
\end{equation}
where $\omega_{Bog}$ is the frequency of a Bogoliubov phonon. Faraday waves are formed whenever $a = n^2$, that is, when the condition $2\hbar\omega_{Bog} = n\hbar\Omega$ is met. This can be seen as the FW forming whenever the energy associated to two Bogoliubov phonons is an integer multiple of the excitation energy $\hbar\Omega$. This is in agreement with the interpretation presented in \cite{Capuzzi, Kagan07}, where Faraday waves are seen as two counter propagating phonons.

This, in fact, reinforces the claim presented in the previous section in which the phase velocity of the Faraday waves corresponds to the superfluid speed of sound. Using Eqs.~(\ref{eq:phase}) and (\ref{eq:kFW}), we obtain an expression for the phase velocity of the $n$-th Faraday excitation as
\begin{equation}\label{eq:phaseFlo}
v_{ph}^{(n)} = \sqrt{\frac{2}{M}}\sqrt{\frac{(\hbar k^{(n)}_{FW})^2}{2M} + E_F}.
\end{equation}

This expression fits our data very well and is in agreement with other theoretical models, as shown by the gray triangles in Fig.~\ref{fig:fig6}, indicating that our simple model offers a good description of the system.

Finally, we further analyze the stability properties of the system. Using the expressions for $a$ and $q$ we can map Figure~\ref{fig:fig3} into an equivalent stability-like diagram and obtain Figure~\ref{fig:fig9}, here we indicate with vertical lines the position of the instability tongues of the Mathieu equation ($a = n^2$ with $n = 1,\,2$ and $3$). We see that the regions where FW are observed (darker regions) correspond to the unstable solutions of the Mathieu equation.

This analysis introduces an interesting way to visualize the competition between the driving and the breathing frequencies in the formation of FW, and provides a universal tool to determine under which conditions Faraday waves can be formed regardless of the interaction strength.

\section{Conclusions}

In this work we have observed and characterized Faraday waves in strongly interacting superfluids composed by atomic pairs of $^6$Li atoms with tunable interactions. By measuring the Faraday wavevector of the generated pattern as a function of the interaction parameter we probe the excitation spectrum of the system, providing a novel way to measure microscopic quantities such as the speed of sound in different superfluid regimes. Our data shows good agreement with mean-field numerical simulations. We have also presented an analytic model that provides a good qualitative description of the phenomenon in terms of stability analysis, allowing the exploration of the parameters space. 

In the near future we plan to explore alternative excitation schemes to produce the Faraday patterns in the BEC-BCS crossover. We also plan to implement momentum-resolved imaging techniques to address additional details in the study of this collective excitations.

\section{Acknowledgments}

We would like to thank Giacomo Roati (LENS \& INO-CNR), Rocío Jáuregui Renaud, Víctor Romero-Rochín, Rosario Paredes Gutiérrez, and Santiago Caballero Benítez (IFUNAM) for fruitful discussion. In particular, we thank Víctor Romero for providing useful ideas in the development of the theory presented in section~\ref{sec:floquet}. We also thank the technical support from Carlos Alberto Gardea Flores, Rodrigo Alejandro Gutiérrez Arenas and Maira Pérez Vielma from the electronics workshop at IFUNAM, as well as Roberto Gleason Villagrán for his support in the development of our research infrastructure.

We acknowledge the following grants: Instituto de Física UNAM (PIIF-8 and PIIF-9); DGAPA-UNAM (PAPIIT projects IA101716, IN103818, IN109021, and IN109619); CONACyT (Ciencia Básica 255573, 254942, and A1-S-39242; Laboratorio Nacional 299057, 314850, and 315838), and Coordinación de la Investigación Científica UNAM (grant number LANMAC-2019).

D.H.R., J.E.P.C., A.R.L. and A.G.V acknowledge their scholarships from CONACyT (programs 000306 and  000328, and Ciencia Básica 254942), and DGAPA-UNAM (PAPIIT projects IA101716, and IN103818).

We finally want to thank the company Seman Baker S.A. de C.V. for their generous donation of numerous machined pieces.

\end{document}